\documentclass[a4paper,12pt]{article}
\usepackage[utf8]{inputenc}

\usepackage{amsmath}
\usepackage{thmtools}
\usepackage{mathpartir}
\usepackage{amsthm}

\usepackage{xparse}
\usepackage{xspace}

\usepackage{subfiles}
\usepackage{caption}
\usepackage{subcaption}
\usepackage{hyperref}

\usepackage{separationLogic}
\usepackage{collections}
\usepackage{genMath}
\usepackage{tuples}
\usepackage{misc}
\usepackage{contextIDs}
\usepackage{resources}
\usepackage{notes}
\usepackage{evalExps}
\usepackage{evalContexts}
\usepackage{functions}
\usepackage{substitution}
\usepackage{algebra}

\newtheorem{theorem}{Theorem}[section]
\newtheorem{definition}[theorem]{Definition}
\newtheorem{lemma}[theorem]{Lemma}
\newtheorem{notation}[theorem]{Notation}{\bfseries}{\itshape}
{\bfseries}{\itshape}
\newtheorem{corollary}[theorem]{Corollary}
\newtheorem{assumption}[theorem]{Assumption}
\newtheorem{example}[theorem]{Example}
\newtheorem{remark}[theorem]{Remark}
\newtheorem{workflow}[theorem]{Workflow}

	\RenewDocumentCommand{\revRemark}{o m}{}
	\RenewDocumentCommand{\revQuestion}{o m}{}
	\RenewDocumentCommand{\TodoTextNote}{o m}{}

\NewDocumentCommand{\deleteBlock}{m}{}

\begin{document}

\author{
	Tobias Reinhard\\
	imec-DistriNet Research Group, KU Leuven\\
	tobias.reinhard@kuleuven.be
}
\title{
	Completeness Thresholds for Memory Safety of Array Traversing Programs:\\ 
	Early Technical Report
}
\date{\today}

\maketitle

\begin{abstract}
	In this early technical report on an ongoing project, we present -- to the best of our knowledge -- the first study of completeness thresholds for memory safety proofs.
	Specifically we consider heap-manipulating programs that iterate over arrays without allocating or freeing memory.
	We present the first notion of completeness thresholds for program verification which reduce unbounded memory safety proofs to bounded ones.
	Moreover, we present some preliminary ideas on how completeness thresholds can be computed for concrete programs.
\end{abstract}

\tableofcontents





\NewDocumentCommand{\assSet}{}{\metaVar{A}}
\NewDocumentCommand{\indexVar}{}{\metaVar{i}}
\NewDocumentCommand{\indexSetVar}{}{\metaVar{I}}
\NewDocumentCommand{\assFam}{o}
	{\ensuremath{
		\assVar
		\IfValueT{#1}{(#1)}
	}\xspace}

\NewDocumentCommand{\preVar}{}{\metaVar{M}}

\NewDocumentCommand{\HeapPredSet}{}{\metaVar{P}}
\NewDocumentCommand{\hPredVar}{}{\metaVar{p}}

\NewDocumentCommand{\setVar}{m}{\ensuremath{\mathcal{#1}}\xspace}

\NewDocumentCommand{\HeapLocSet}{}{\setVar{L}}
\NewDocumentCommand{\VarSet}{}{\setVar{X}}
\NewDocumentCommand{\dataValVarSet}{}{\setVar{X_D}}
\NewDocumentCommand{\ptrVarVarSet}{}{\setVar{X_L}}
\NewDocumentCommand{\ProcNameSet}{}{\setVar{X_P}}
\NewDocumentCommand{\ValueSet}{}{\setVar{V}}
\NewDocumentCommand{\OpSet}{}{\metaVar{Ops}}
\NewDocumentCommand{\BaseTypeSet}{}{\metaVar{BaseTypes}}
\NewDocumentCommand{\ComplexTypeSet}{}{\metaVar{ComplexTypes}}
\NewDocumentCommand{\TypeSet}{}{\metaVar{Types}}
\NewDocumentCommand{\DataTypeSet}{}{\metaVar{DataTypes}}
\NewDocumentCommand{\DataSet}{}{\metaVar{Data}}

\NewDocumentCommand{\ExpSet}{}{\metaVar{Exps}}
\NewDocumentCommand{\LocExpSet}{}{\metaVar{LocExps}}
\NewDocumentCommand{\DataExpSet}{}{\metaVar{DataExps}}
\NewDocumentCommand{\CmdSet}{}{\metaVar{Cmds}}
\NewDocumentCommand{\BasicCmdSet}{}{\metaVar{BCmds}}
\NewDocumentCommand{\HeapCmdSet}{}{\metaVar{HCmds}}
\NewDocumentCommand{\CmdPotErrSet}{}{\metaVar{Cmds^{+}}}
\NewDocumentCommand{\ProcSet}{}{\metaVar{Procs}}
\NewDocumentCommand{\TransUnitSet}{}{\metaVar{TUs}}


\NewDocumentCommand{\hlocVar}{}{\metaVar{l}}
\NewDocumentCommand{\nVar}{}{\metaVar{n}}
\NewDocumentCommand{\nextNVar}{}{\metaVar{n'}}
\NewDocumentCommand{\zVar}{}{\metaVar{z}}
\NewDocumentCommand{\bVar}{}{\metaVar{b}}
\NewDocumentCommand{\varVar}{}{\metaVar{x}}
\NewDocumentCommand{\freshVarVar}{}{\metaVar{x_{\text{fresh}}}}
\NewDocumentCommand{\dataVarVar}{}{\metaVar{d}}
\NewDocumentCommand{\ptrVarVar}{}{\metaVar{ptr}}
\NewDocumentCommand{\procNameVar}{}{\metaVar{p}}
\NewDocumentCommand{\valVar}{}{\metaVar{v}}
\NewDocumentCommand{\dataValVar}{o}
	{\metaVar{
		v^d
		\IfValueT{#1}{_{#1}}
	}}
\NewDocumentCommand{\nextDataVar}{o}
	{\metaVar{
		v^{d\,\prime}
		\IfValueT{#1}{_{#1}}
	}}
\NewDocumentCommand{\nextValVar}{}{\metaVar{v^\prime}}
\NewDocumentCommand{\expVar}{}{\metaVar{e}}
\NewDocumentCommand{\nextExpVar}{}{\metaVar{e'}}
\NewDocumentCommand{\locExpVar}{}{\metaVar{e^L}}
\NewDocumentCommand{\nextLocExpVar}{}{\metaVar{e^{L\prime}}}
\NewDocumentCommand{\dataExpVar}{o}
	{\metaVar{
			e^d
			\IfValueT{#1}{_{#1}}
	}}
\NewDocumentCommand{\nextDataExpVar}{}{\metaVar{e^{d\prime}}}
\NewDocumentCommand{\cmdVar}{}{\metaVar{c}}
\NewDocumentCommand{\nextCmdVar}{}{\metaVar{c'}}
\NewDocumentCommand{\cmdPotErr}{}{\metaVar{\cmdVar^{+}}}
\NewDocumentCommand{\hCmdVar}{}{\metaVar{hc}}
\NewDocumentCommand{\bCmdVar}{}{\metaVar{bc}}
\NewDocumentCommand{\fctVar}{}{\metaVar{f}}
\NewDocumentCommand{\opVar}{}{\metaVar{op}}
\NewDocumentCommand{\typeVar}{}{\metaVar{T}}
\NewDocumentCommand{\nextTypeVar}{}{\metaVar{T^\prime}}
\NewDocumentCommand{\dataTypeVar}{}{\metaVar{D}}
\NewDocumentCommand{\nextDataTypeVar}{}{\metaVar{D^\prime}}
\NewDocumentCommand{\procVar}{}{\metaVar{proc}}
\NewDocumentCommand{\tuVar}{}{\metaVar{tu}}
\NewDocumentCommand{\nextTuVar}{}{\metaVar{tu'}}

\NewDocumentCommand{\nSeqVar}{}{\metaVar{s}}
\NewDocumentCommand{\rangeVar}{}{\metaVar{R}}

\NewDocumentCommand{\arrayVar}{}{\metaVar{a}}
\NewDocumentCommand{\listVar}{}{\metaVar{li}}
\NewDocumentCommand{\sizeVar}{}{\metaVar{s}}

\NewDocumentCommand{\UnitType}{}{\fixedFuncNameFont{Unit}}
\NewDocumentCommand{\PtrType}{O{\Z}}
	{\ensuremath{
			{}^\rightarrow
			#1
	}\xspace}

\NewDocumentCommand{\opPlusLocsZ}{}{\ensuremath{\operatorname{+_{\HeapLocSet\Z}}}}

\NewDocumentCommand{\memoryError}{}
	{\ensuremath{\fixedPredNameFont{error}}\xspace}

\NewDocumentCommand{\unitVal}{}
	{\ensuremath{\fixedPredNameFont{()}}\xspace}


\NewDocumentCommand{\rangeFromTo}{m m}
	{\ensuremath{
		[#1\ \keyword{to}\ #2]
	}}

\NewDocumentCommand{\cmdReadHloc}{m}
{\ensuremath{
		!#1
	}\xspace}

\NewDocumentCommand{\cmdAssignHloc}{m m}
	{\ensuremath{
		\cmdReadHloc{#1} := #2
	}}

\NewDocumentCommand{\cmdLet}{m m m}
    {\ensuremath{
        \keyword{let}\ #1 := #2\ \keyword{in}\ #3
    }}

\NewDocumentCommand{\cmdIf}{m m m}
    {\ensuremath{
        \keyword{if}\ #1\ \keyword{then}\ #2\ \keyword{else}\ #3
    }}

\NewDocumentCommand{\cmdIfNoElse}{m m}
{\ensuremath{
		\keyword{if}\ #1\ \keyword{then}\ #2
}}

\NewDocumentCommand{\cmdWhile}{m m}
{\ensuremath{
		\keyword{while}\ #1\ \keyword{do}\ #2
}}

\NewDocumentCommand{\cmdWhileInv}{m m m}
{\ensuremath{
		\keyword{while}\ #1\ \keyword{inv}\ #2\ \keyword{do}\ #3
}}

\NewDocumentCommand{\cmdFor}{m m m}
{\ensuremath{
		\keyword{for}\ #1\ \keyword{in}\ #2\ \keyword{do}\ #3
}}

\NewDocumentCommand{\cmdForInv}{m m m m}
{\ensuremath{
		\keyword{for}\ #1\ \keyword{in}\ #2\ \keyword{inv}\ #3\ \keyword{do}\ #4
}}

\NewDocumentCommand{\cmdContextID}{}
{\contextIdFont{cmds}}

\NewDocumentCommand{\cmdGetProcNames}{o}
{\ensuremath{
		\fixedFuncNameFont{procNames}_\cmdContextID
		\IfValueT{#1}{(#1)}
	}\xspace}


\NewDocumentCommand{\heapCmdContextID}{}
	{\contextIdFont{hcmds}}


\NewDocumentCommand{\heapPredContextID}{}
	{\contextIdFont{hpreds}}

\NewDocumentCommand{\cmdLetProc}{m m}
{\ensuremath{
		\keyword{letproc}\ #1\ \{ #2 \}
}}

\NewDocumentCommand{\varDecl}{m m}
{\ensuremath{
		#1 : #2
}}

\NewDocumentCommand{\mainFct}{}{\keyword{main}}


\NewDocumentCommand{\safeCmdHeapName}{}
	{\ensuremath{\fixedPredNameFont{safe}}}

\NewDocumentCommand{\safeCmdHeap}{m m}
	{\ensuremath{
			\safeCmdHeapName(#1, #2)
	}}

\NewDocumentCommand{\safeCmdAssName}{}
{\ensuremath{\fixedPredNameFont{safe}_\assertionContextID}}

\NewDocumentCommand{\safeCmdAss}{m m}
{\ensuremath{
		\safeCmdAssName(#1, #2)
}}

\NewDocumentCommand{\TypingCtxtSet}{}
	{\ensuremath{TypingCtxts}\xspace}
\NewDocumentCommand{\typingCtxtVar}{}
	{\metaVar{\Gamma}\xspace}

\NewDocumentCommand{\ProcCtxtSet}{}{\ensuremath{\metaVar{ProcCtxts}}\xspace}
\NewDocumentCommand{\procCtxtVar}{}{\metaVar{\Lambda}\xspace}
\NewDocumentCommand{\nextProcCtxtVar}{}{\metaVar{\Lambda'}\xspace}

\NewDocumentCommand{\procCtxtContextID}{}
	{\contextIdFont{procCtxt}}

\NewDocumentCommand{\procCtxtGetProcNames}{o}
	{\ensuremath{
		\fixedFuncNameFont{procNames}_\procCtxtContextID
		\IfValueT{#1}{(#1)}
	}\xspace}

\NewDocumentCommand{\SymbLocSet}{}{\metaVar{\HeapLocSet^{\contextIdFont{symb}}}}
\NewDocumentCommand{\slocVar}{}{\metaVar{L}}

\NewDocumentCommand{\SymbValueSet}{}{\metaVar{\ValueSet^\contextIdFont{symb}}}
\NewDocumentCommand{\svalVar}{}{\metaVar{\valVar^\contextIdFont{s}}}
\NewDocumentCommand{\nextSvalVar}{}{\metaVar{\valVar^{\contextIdFont{s}\,\prime}}}

\NewDocumentCommand{\preProgTuple}{m m}
{\ensuremath{
		\{ #1 \}\ #2
	}\xspace}

\section{Introduction}

In this early technical report on an ongoing project, we present -- to the best of our knowledge -- the first study of completeness thresholds for memory safety proofs.
Specifically we consider heap-manipulating programs that iterate over arrays without allocating or freeing memory.
We present the first notion of completeness thresholds for program verification which reduce unbounded memory safety proofs to bounded ones.
Moreover, we present some preliminary ideas on how completeness thresholds can be computed for concrete programs.

\paragraph{Unbounded vs Bounded Proofs}
Memory safety is a very basic property we want to hold for every critical program, regardless of its nature or purpose.
Yet, it remains hard to prove and in general requires us to write tedious, inductive proofs.
One way to automate the verification process is to settle on bounded proofs and accept bounded guarantees.

Consider a program \cmdVar that searches through an array of size \sizeVar.
An unbounded memory safety proof for \cmdVar would yield that the program is safe for any possible input, in particular for any array size, i.e., 
$\forall \sizeVar.\ \texttt{safe}(\cmdVar)$.
A bounded proof that only considers input sizes smaller than 10 would only guarantees that the program is safe for any such bounded array, i.e., 
$\forall \sizeVar < 10.\ \texttt{safe}(\cmdVar)$.

\paragraph{Completeness Thresholds}
Approximating unbounded proofs by bounded ones is a technique often used in model checking.
Hence, the relationship between bounded and unbounded proofs about finite state transition systems has been studied extensively~\cite{Biere1999SymbolicMC, Clarke2004CompletenessAC, Kroening2003EfficientCO, Bundala2012OnTM, Abdulaziz2018FormallyVA, Heljanko2005IncrementalAC, Awedh2004ProvingMP, McMillan2003InterpolationAS}.
For a finite transition system $T$ and a property of interest $\phi$, a \emph{completeness threshold} is any number $k$ such that we can prove $\phi$ by only examining path prefixes of length $k$ in $T$, i.e., $T \models_k \phi \Rightarrow T \models \phi$~\cite{Clarke2004CompletenessAC}~\footnote{
	Note that the term completeness threshold is used inconsistently in literature.
	Some papers such as~\cite{Clarke2004CompletenessAC} use the definition above, according to which completeness thresholds are not unique.
	Others such as~\cite{Kroening2003EfficientCO} define them as the minimal number $k$ such that $T \models_k \phi \Rightarrow T \models \phi$, which makes them unique.
}.
Over the years, various works characterised over-approximations of least completeness thresholds for different types of properties~$\phi$.
These over-approximations are typically described in terms of key attributes of the transition system $T$, such as the recurrence diameter~\cite{Kroening2003EfficientCO}.

Heap-manipulating programs are essentially infinite state transition systems.
Hence, in general, these key attributes are infinite.
This vast structural difference between the programs we are interested in and the transition systems for which completeness thresholds have been studied prevents us from reusing any of the existing definitions or results.

In \S~\ref{sec:NotationsAndDefinitions} we start by presenting basic definitions and notations that are used throughout this work.
In \S~\ref{sec:Syntax} and \S~\ref{sec:DynamicSemantics} we present the syntax and semantics of the programming language we consider.
In \S~\ref{sec:Assertions} and \S~\ref{sec:MemorySafety} we formalise our assertion language as well as the notion of memory safety we consider.
Our study of completeness thresholds relies on verification conditions, which we define in \S~\ref{sec:VerificationConditions}.
We introduce completeness thresholds and study their properties in \S~\ref{sec:CompletenessThresholds}.
Finally, we describe our envisioned roadmap for this work in progress in \S~\ref{sec:Roadmap}.

\section{General Notation and Basic Definitions}\label{sec:NotationsAndDefinitions}

The following definitions and notations will be used throughout this work.

\begin{definition}[Tuples]
	For any set $X$ we denote the set of tuples over $X$ as 
	$$
		X^* 
		\ := \
		\bigcup_{n\in\N}\ \setOf[(x_1, \dots x_n)]{x_1, \dots, x_n \in X}.
	$$
	We denote tuples by overlining the variable name, i.e., $\overline{x} \in X^*$, except if it is clear from the context.
	
	For any set $X$ and for any tuple $\overline{t} = (x_1, \dots, x_n) \in X^*$, we denote the tuple's length by
	$\tupleLen{\overline{t}} = n$.
	
	We define the following notation for appending a single element $e\in X$ to a tuple:
	$$
		\tupleAppendElem{\overline{t}}{e} 
		\ :=\
		(x_1, \dots, x_n, e).
	$$
\end{definition}

\begin{definition}[Non-Empty Tuples]
	For any set $X$ we denote the set of non-empty tuples over $X$ as 
	$$
	X^+
	\ := \
	\setOf[t\in X^*]{\tupleLen{t} > 0}.
	$$
\end{definition}

\begin{definition}[Disjoint Union]
	Let $A, B$ be sets.
	We define their disjoint union as
	$$
		A \disjCup B \ := \ A \cup B
	$$
	if $A \cap B = \emptyset$ and leave it undefined otherwise.
\end{definition}

\NewDocumentCommand{\eqClass}{m o}
{\ensuremath{
		\langle #1 \rangle
		\IfValueT{#2}{_{#2}}
	}\xspace}

\begin{definition}[Equivalence Classes]
	Let $R \subseteq X \times X$ be an equivalence relation.
	For any $x \in X$, we define the notation
	$$
	\eqClass{x}[R] :=
	\setOf[x^\prime \in X]{R(x, x^\prime)}
	\in X/R.
	$$
	We omit $R$ whenever it is clear from the context and write \eqClass{x} instead of \eqClass{x}[R].
\end{definition}

\begin{notation}[Homomorphic \& Isomorphic]
	Let $A, B$ be algebraic structures.
	We define the following notations:
	\begin{itemize}
		\item $A \homomorphic B$ 
			expresses that $A$ and $B$ are homomorphic.
		\item $A \isomorphic B$
			expresses that $A$ and $B$ are isomorphic.
	\end{itemize}
\end{notation}

\begin{remark}[Canonical Homomorphism from Tuples to Sets]
	Let $X$ be a set.
	Then, $X^*$ with concatenation and $\powerSetOf{X}$ with union are both monoids and the canonical homomorphism from $X^*$ to $\powerSetOf{X}$ is
	$
		(x_1, \dots, x_n) \mapsto \setOf{x_1, \dots, x_n}.
	$
\end{remark}

\begin{definition}[Congruence Relation between Tuples and Sets]
	Let $X$ be a set and let $h_X : X^* \rightarrow \powerSetOf{X}$ be the canonical homomorphism.
	We define the congruence relation
	$\congruent_X\ \subseteq 
		(X^* \times \powerSetOf{X})
		\cup
		(\powerSetOf{X} \times X^*)
	$
	such that the following holds for all 
	$\overline{t} \in X^*$ and all $S \in \powerSetOf{X}$:
	$$
		S \congruent_X \overline{t}
		\quad\Leftrightarrow\quad
		\overline{t} \congruent_X S
		\quad\Leftrightarrow\quad
		h_X(\overline{t}) = S
	$$
	Whenever the base set $X$ is clear from the context, we write \congruent instead of $\congruent_X$.

\end{definition}


\NewDocumentCommand{\hcmdInterpret}{o}
{\ensuremath{
		\mathcal{I}_\heapCmdContextID
		\IfValueT{#1}{(#1)}
	}\xspace}

\NewDocumentCommand{\hpredInterpret}{o}
	{\ensuremath{
		\mathcal{I}_\heapPredContextID
		\IfValueT{#1}{(#1)}
	}\xspace}

\newpage
\section{Syntax}\label{sec:Syntax}

In this section we define the syntax of the programming language that we use in the rest of this work.

\begin{definition}[Variables]
	We define \VarSet to be an infinite set of variable symbols.
\end{definition}

Our language allows for simple pointer arithmetic of the form $\hlocVar \opPlusLocsZ \zVar$ where \hlocVar is a heap location and \zVar is an offset.

\NewDocumentCommand{\HeapObjSet}{}{\metaVar{HObjs}}
\NewDocumentCommand{\hObjVar}{}{\metaVar{o}}
\NewDocumentCommand{\HeapIdxSet}{}{\metaVar{HIdxs}}
\NewDocumentCommand{\hIdxVar}{}{\metaVar{idx}}

\begin{definition}[Structured Heap Locations]
	We define the sets of heap objects \HeapObjSet and heap indices \HeapIdxSet to be infinite sets with $\N \subset \HeapIdxSet$.
	Further, we define the set of heap locations as
	$$
		\HeapLocSet \quad := \quad \HeapObjSet \times \HeapIdxSet.
	$$
	We denote heap locations by \hlocVar, heap objects by \hObjVar and heap indices by \hIdxVar.
\end{definition}

\begin{definition}[Unit Type]
	We define the unit type as $\UnitType := \setOf{\unitVal}$.
\end{definition}

\begin{definition}[Types]
	We define the set of types \TypeSet syntactically as follows:
	$$
	\begin{array}{l c l}
	\typeVar \in \TypeSet &:=
	&\HeapLocSet \alt \HeapObjSet \alt \Z \alt \B \alt \UnitType
	\end{array}
	$$
\end{definition}

Operations are pure functions that map inputs to output and cannot access the heap.

\begin{definition}[Operations]
	We define \OpSet to be a set of operations with 
	$\OpSet \subseteq \setOf[f: D\rightarrow C]{ D\in\TypeSet^*, C\in\TypeSet}$
	and with $<,\, + \in \OpSet$.
	
	For each $f: D \rightarrow C \in \OpSet$, we use the following notation
	$\dom{f} = D$ and
	$\codom{f} = C$.
\end{definition}

While this work currently only deals with arrays, our plan is to investigate completeness thresholds for programs that deal with arbitrary tree-like inductive data structures.
In order to keep the semantics of our language modular and to allow for easy extensions, we introduce an unspecified set of \emph{heap commands} that captures the APIs of the data structures we are interested in.

\begin{definition}[Heap Commands]
	We define \HeapCmdSet to be a set of symbols.
	Every $\hCmdVar \in \HeapCmdSet$ represents a command that accesses the heap.
\end{definition}

\begin{definition}[Program syntax]
	We define the set of commands \CmdSet, expressions \ExpSet and values \ValueSet syntactically by the grammar presented in Fig.~\ref{fig:def:ProgSyntax}.
\end{definition}

\begin{figure}
	\begin{subfigure}{\textwidth}
		$$
		\begin{array}{lcl p{1cm}l}
			\hlocVar \in \HeapLocSet 
				&&&&\text{Heap locations}
			\\
			\hObjVar \in \HeapObjSet
				&&&&\text{Heap Objects}
			\\
			\zVar \in \Z
			\\
			\bVar \in \B
			\\
			\varVar  \in  \VarSet
			&&&&\text{Variables}
			\\
			\opVar \in \OpSet &
				&
				&&\text{Primitive Operations}
			\\
			\hCmdVar \in \HeapCmdSet
				&&&&\text{Heap Commands}
			\\
			\\
			\valVar \in \ValueSet &::=
				&\hlocVar \alt \hObjVar \alt \zVar \alt \bVar \alt \unitVal
				&&\text{Values}
			\\
			\expVar \in \ExpSet &::=
				&\valVar \alt  
					\varVar \alt
					\opVar(\overline{\expVar})
				&&\text{Expressions}
			\\
			\\
			\cmdVar \in \CmdSet &::= 
				&\expVar \alt 
				\cmdLet{\varVar}{\cmdVar}{\cmdVar} \alt
				\\
				&&\cmdIf{\expVar}{\cmdVar}{\cmdVar} \alt
				\\
				&&\cmdWhile{\cmdReadHloc{\expVar}}{\cmdVar} \alt
				\\
				&&\cmdFor{\varVar}{\rangeFromTo{\expVar}{\expVar}}{\cmdVar} \alt
				\\
				&&\cmdReadHloc{\expVar} \alt 
				\cmdAssignHloc{\expVar}{\expVar} \alt
				\hCmdVar(\overline{\expVar})
				&&\text{Heap access}
		\end{array}
		$$
		\caption{Values, expressions and commands.}
	\end{subfigure}
	
	\begin{subfigure}{\textwidth}
		$$
		\begin{array}{c l l l}
			\cmdVar ; \nextCmdVar 
			&\ :=\
			&\cmdLet{\varVar}{\cmdVar}{\nextCmdVar}
			&\text{where \varVar is not free in \nextCmdVar}
			\\
			\cmdIfNoElse{\expVar}{\cmdVar}
			&\ :=\
			&
			\cmdIf{\expVar}{\cmdVar}{\unitVal}
		\end{array}
		$$
		\caption{Syntactic sugar.}
	\end{subfigure}

	\caption{Program syntax.}
	\label{fig:def:ProgSyntax}
\end{figure}

\newpage
\newpage
\section{Dynamic Semantics}\label{sec:DynamicSemantics}

\NewDocumentCommand{\MemTraceSet}{}{\metaVar{MemTraces}}
\NewDocumentCommand{\cmdMemTrace}{o}
{\ensuremath{
		\fixedFuncNameFont{memTrace}_\cmdContextID
		\IfValueT{#1}{(#1)}
	}\xspace}

\NewDocumentCommand{\memTraceVar}{}{\metaVar{tr}}
\NewDocumentCommand{\nextMemTraceVar}{o}
{\metaVar{
		tr^\prime
		\IfValueT{#1}{_{#1}}
}}

\NewDocumentCommand{\progRedStepSymb}{}
	{\ensuremath{\rightsquigarrow_\contextIdFont{prog}}\xspace}

\NewDocumentCommand{\progRedStep}{m m m m m m}
	{\ensuremath{
		#1, #2, #3
		\tuRedStepSymb
		#4, #5, #6
	}\xspace}


\NewDocumentCommand{\tuRedStepSymb}{}
	{\ensuremath{\rightsquigarrow_\contextIdFont{tu}}\xspace}
	
\NewDocumentCommand{\tuRedStep}{m m m m}
	{\ensuremath{
			#1, #2 
			\tuRedStepSymb
			#3, #4
		}\xspace}

\NewDocumentCommand{\cmdRedStepSymb}{}
	{\ensuremath{\rightsquigarrow_\contextIdFont{cmd}}\xspace}
	
\NewDocumentCommand{\cmdRedStepStarSymb}{}
	{\ensuremath{\rightsquigarrow_\contextIdFont{cmd}^*}\xspace}

\NewDocumentCommand{\cmdRedStep}
	{O{\pheapVar} m O{\pheapVar} m}
	{\ensuremath{
		#1, #2
		\cmdRedStepSymb
		#3, #4
	}\xspace}

\NewDocumentCommand{\cmdRedStepStar}
{O{\pheapVar} m O{\pheapVar} m}
{\ensuremath{
		#1, #2,
		\cmdRedStepStarSymb
		#3, #4
	}\xspace}

In order to keep things simple, our language uses a heap but no store.
Hence, variables are actually constants that can be bound to values via \keyword{let} commands.
As a consequence, the entire evaluation state of a program is represented by the heap and the program itself.
Further, expressions are pure, hence their evaluation does not depend on the heap.

\begin{definition}[Evaluation of Closed Expressions]\label{def:EvaluationOfClosedExpressions}
	We define a partial evaluation function $\evalExp{\cdot}: \ExpSet \partialFunArrow \ValueSet$ on expressions by recursion on the structure of expressions as follows:
	$$
	\begin{array}{l l l l}
		\evalExp{\valVar} &:= &\valVar 
			&\text{if}\quad \valVar \in \ValueSet,
		\\
		\evalExp{\opVar(\expVar_1, \dots, \expVar_n)} 
			&:=
			&\opVar(\evalExp{\expVar_1}, \dots, \evalExp{\expVar_n})
			&\text{if}\quad \bot\not\in\setOf{\evalExp{\expVar_1}, \dots, \evalExp{\expVar_n}}
		\\
			&&&\text{and}\quad (\evalExp{\expVar_1}, \dots, \evalExp{\expVar_n}) \in \dom{\opVar},
		\\
		\evalExp{\expVar} &:= &\bot &\text{otherwise}.
	\end{array}
	$$
	We identify closed expressions \expVar with their ascribed value \evalExp{\expVar}.
\end{definition}

\begin{definition}[Evaluation Context]
	We define the set of evaluation contexts \EvalCtxtSet syntactically as follows:
	$$
	\begin{array}{l c l}
		\evalCtxt \in \EvalCtxtSet &\ ::=\
				\cmdLet{\varVar}{\evalCtxtHole}{\cmdVar} 
	\end{array}
	$$
	For any $\cmdVar \in \CmdSet$ and $\evalCtxt \in \EvalCtxtSet$, we define 
	$\evalCtxt[\cmdVar] := \subst{\evalCtxt}{\evalCtxtHole}{\cmdVar}$.
\end{definition}

\NewDocumentCommand{\freeVars}{o}
{\ensuremath{
		\fixedFuncNameFont{freeVars}
		\IfValueT{#1}{(#1)}
}\xspace}

\begin{definition}[Free Variables (Commands)]\label{def:freeVariables_commands}
	We define free variables in the usual way.
	For any command \cmdVar we denote the set of variables that occur freely in \cmdVar by \freeVars[\cmdVar].
\end{definition}

\begin{definition}[Substitution (Commands)]\label{def:substitution_commands}
	We define substitution in the usual way.
	For any command \cmdVar, variable \varVar and expression \expVar, we denote the result of substituting every free occurrence of \varVar in \cmdVar with \expVar by \subst{\cmdVar}{\varVar}{\expVar}.
	Further, we extend substitutions to tuples of variables and expressions in the canonical way.
\end{definition}

We explicitly model memory errors in our operational semantics.
This way we know that 
(i)~any execution which ends in a value does not involve memory errors and 
(ii)~any execution that does involve memory errors ends in the dedicated error state \memoryError.

\begin{definition}[Memory Errors]
	We denote the memory error state by \memoryError and the set of potentially erroneous commands by
	$\CmdPotErrSet := \CmdSet \cup \setOf{\memoryError}$.
	We denote potentially erroneous commands by \cmdPotErr.
\end{definition}

Heaps are finite collections of resources that can be manipulated by commands.

\begin{definition}[Physical Resources]
	We define the set of physical resources \PhysResSet syntactically as follows:
	$$
	\begin{array}{l c l}
		\presVar \in \PhysResSet 
		&\ ::=\
			&\slPointsTo{\hlocVar}{\valVar}
			\end{array}
	$$
	$$
	\begin{array}{c p{0.5cm} c}
		\hlocVar \in \HeapLocSet
		&&\valVar \in \ValueSet
	\end{array}
	$$
\end{definition}

\begin{definition}[Physical Heaps]
	We define the set of physical heaps as 
	$$\PhysHeapSet\ :=\ \finPowerSetOf{\PhysResSet}$$
	                and the function 
	$\phGetLocs: \PhysHeapSet \rightarrow \finPowerSetOf{\HeapLocSet}$
	mapping physical heaps to the sets of allocated heap locations as
	$$
	\begin{array}{l c l}
		\phGetLocs(\pheapVar) 
		& :=
		&\setOf
			[\hlocVar \in \HeapLocSet]
			{
				\exists \valVar \in \ValueSet.\
				\slPointsTo{\hlocVar}{\valVar} \in \pheapVar
			}.
	\end{array}
	$$
	We denote physical heaps by \pheapVar.
\end{definition}

\begin{definition}[Basic Commands]
	We define the set of basic commands \BasicCmdSet syntactically as follows:
	$$
	\begin{array}{l l l}
		\bCmdVar \in \BasicCmdSet &::= 
			&\expVar \alt 
				\cmdLet{\varVar}{\bCmdVar}{\bCmdVar} \alt
		\\
			&&\cmdIf{\expVar}{\bCmdVar}{\bCmdVar} \alt
		\\
			&&\cmdWhile{\cmdReadHloc{\expVar}}{\bCmdVar} \alt
		\\
			&&\cmdFor{\varVar}{\rangeFromTo{\expVar}{\expVar}}{\bCmdVar} \alt
		\\
			&&\cmdReadHloc{\expVar} \alt 
				\cmdAssignHloc{\expVar}{\expVar}
	\end{array}
	$$
\end{definition}

The set of basic commands is the subset of \CmdSet that consists exactly of those commands that do not involve any heap command call $\hCmdVar(\overline{\expVar})$.
Remember that the set of heap commands captures the APIs of data structures.
We do not want to change our operational semantics each time we want to consider a new data structure.
Hence, we assume that there exists an interpretation for each heap command that describes its behaviour in terms of basic commands.

\begin{assumption}[Heap Command Interpretation]
	We assume that there exists a function
	$\hcmdInterpret : \HeapCmdSet \rightarrow (\BasicCmdSet \times \VarSet^*)$
	that maps each heap command to a basic command and a vector of variables.
	Further, for every mapping of the form
	$\hcmdInterpret[\hCmdVar] = (\bCmdVar, (\varVar_1, \dots, \varVar_n))$ the following two properties hold:
	\begin{itemize}
		\item $\displaystyle \bigwedge_{i \neq j} \varVar_i \neq \varVar_j$
		\item $\freeVars[\bCmdVar] \subseteq \setOf{\varVar_1, \dots, \varVar_n}$.
	\end{itemize}
\end{assumption}

\NewDocumentCommand{\cmdRedRuleName}{m}{CmdRed-#1}

\NewDocumentCommand{\cmdRedEvalCtxtName}{}{\cmdRedRuleName{EvalCtxt}}
\NewDocumentCommand{\cmdRedEvalCtxt}{}
{
	\inferrule[\cmdRedEvalCtxtName]
	{\cmdRedStep
		{\cmdVar}
		[\nextPheapVar]{\nextCmdVar}
	}
	{\cmdRedStep
		{\evalCtxt[\cmdVar]}
		[\nextPheapVar]{\evalCtxt[\nextCmdVar]}
	}
}

\NewDocumentCommand{\cmdRedEvalCtxtFailName}{}{\cmdRedRuleName{EvalCtxt-Fail}}
\NewDocumentCommand{\cmdRedEvalCtxtFail}{}
{
	\inferrule[\cmdRedEvalCtxtFailName]
	{\cmdRedStep
		{\cmdVar}
		[\nextPheapVar]{\memoryError}
	}
	{\cmdRedStep
		{\evalCtxt[\cmdVar]}
		[\nextPheapVar]{\memoryError}
	}
}

\NewDocumentCommand{\cmdRedIfTrueName}{}{\cmdRedRuleName{IfTrue}}
\NewDocumentCommand{\cmdRedIfTrue}{}
{
	\inferrule[\cmdRedIfTrueName]
	{}
	{\cmdRedStep
		{\cmdIf{\slTrue}{\cmdVar_t}{\cmdVar_f}}
		{\cmdVar_t}
	}
}

\NewDocumentCommand{\cmdRedIfFalseName}{}{\cmdRedRuleName{IfFalse}}
\NewDocumentCommand{\cmdRedIfFalse}{}
{
	\inferrule[\cmdRedIfFalseName]
	{}
	{\cmdRedStep
		{\cmdIf{\slFalse}{\cmdVar_t}{\cmdVar_f}}
		{\cmdVar_f}
	}
}

\NewDocumentCommand{\cmdRedWhileName}{}{\cmdRedRuleName{While}}
\NewDocumentCommand{\cmdRedWhile}{}
{
	\inferrule[\cmdRedWhileName]
	{\varVar \not\in \freeVars[\cmdVar]}
	{\cmdRedStep
		{\cmdWhile{\cmdReadHloc{\hlocVar}}{\cmdVar}}
		{
			\cmdLet
				{\varVar}
				{\cmdReadHloc{\hlocVar}}
				{\cmdIfNoElse
					{\varVar}
					{(
						\cmdVar;
						\cmdWhile{\cmdReadHloc{\hlocVar}}{\cmdVar}
					)}
				}
		}
	}
}

\NewDocumentCommand{\cmdRedForName}{}{\cmdRedRuleName{For}}
\NewDocumentCommand{\cmdRedFor}{}
{
	\inferrule[\cmdRedForName]
	{}
	{\cmdRedStep
		{\cmdFor{\varVar}{\rangeFromTo{\nVar}{\nextNVar}}{\cmdVar}}
		{
			\cmdIfNoElse{\nVar \leq \nextNVar}
				{(
					\cmdVar;
					\cmdFor{\varVar}{\rangeFromTo{\nVar+1}{\nextNVar}}{\cmdVar}
				)}
		}
	}
}

\NewDocumentCommand{\cmdRedLetName}{}{\cmdRedRuleName{Let}}
\NewDocumentCommand{\cmdRedLet}{}
{
	\inferrule[\cmdRedLetName]
	{}
	{\cmdRedStep
		{\cmdLet{\varVar}{\valVar}{\cmdVar}}
		{\subst{\cmdVar}{\varVar}{\valVar}}
	}
}

\NewDocumentCommand{\cmdRedReadHeapLocName}{}{\cmdRedRuleName{ReadHeapLoc}}
\NewDocumentCommand{\cmdRedReadHeapLoc}{}
{
	\inferrule[\cmdRedReadHeapLocName]
	{
		\slPointsTo{\hlocVar}{\valVar} \in \pheapVar
	}
	{\cmdRedStep
		{\cmdReadHloc{\hlocVar}}
		[\pheapVar]{\valVar}
	}
}

\NewDocumentCommand{\cmdRedReadHeapLocFailName}{}{\cmdRedRuleName{ReadHeapLoc-Fail}}
\NewDocumentCommand{\cmdRedReadHeapLocFail}{}
{
	\inferrule[\cmdRedReadHeapLocFailName]
	{
		\hlocVar \not\in \phGetLocs[\pheapVar]
	}
	{\cmdRedStep
		{\cmdReadHloc{\hlocVar}}
		[\pheapVar]{\memoryError}
	}
}

\NewDocumentCommand{\cmdRedAssignHeapLocName}{}{\cmdRedRuleName{AssignHeapLoc}}
\NewDocumentCommand{\cmdRedAssignHeapLoc}{}
{
	\inferrule[\cmdRedAssignHeapLocName]
	{}
	{\cmdRedStep
		[\phAddResDisj
			{\slPointsTo
				{\hlocVar}
				{\slWildcard}
			}
		]
		{\cmdAssignHloc{\hlocVar}{\valVar}}
		[\phAddResDisj
			{\slPointsTo
				{\hlocVar}
				{\valVar}
			}
		]
		{\unitVal}
	}
}

\NewDocumentCommand{\cmdRedAssignHeapLocFailName}{}{\cmdRedRuleName{AssignHeapLoc-Fail}}
\NewDocumentCommand{\cmdRedAssignHeapLocFail}{}
{
	\inferrule[\cmdRedAssignHeapLocFailName]
	{
		\hlocVar \not\in \phGetLocs[\pheapVar]
	}
	{\cmdRedStep
		{\cmdAssignHloc{\hlocVar}{\valVar}}
		[\pheapVar]{\memoryError}
	}
}

\NewDocumentCommand{\cmdRedDesugarHeapCmdCallName}{}{\cmdRedRuleName{Desugar-HeapCmdCall}}
\NewDocumentCommand{\cmdRedDesugarHeapCmdCall}{}
{
	\inferrule[\cmdRedDesugarHeapCmdCallName]
	{
		\hcmdInterpret[\hCmdVar] = (\bCmdVar, (\varVar_1, \dots, \varVar_n))
	}
	{\cmdRedStep
		{\hCmdVar(\valVar_1, \dots, \valVar_n)}
		{\subst
			{\bCmdVar}
			{(\varVar_1, \dots, \varVar_n)}
			{(\valVar_1, \dots, \valVar_n)}
		}
	}
}

\begin{definition}[Command Reduction Relation]
	We define a command reduction relation \cmdRedStepSymb according to the
	rules presented in Fig.~\ref{fig:def:CmdReductionRules}.
	A reduction step has the form 
	$$\cmdRedStep
		[\pheapVar]{\cmdVar}
		[\nextPheapVar]{\nextCmdVar}.
	$$
	
	We define \cmdRedStepStarSymb as the reflexive transitive closure of \cmdRedStepSymb.
\end{definition}

\begin{figure}
	\begin{mathpar}
		\cmdRedEvalCtxt
		\and
		\cmdRedEvalCtxtFail
		\and
		\cmdRedIfTrue
		\and
		\cmdRedIfFalse
		\and
		\cmdRedWhile
		\and
		\cmdRedFor
		\and
		\cmdRedLet
		\and
		\cmdRedReadHeapLoc
		\and
		\cmdRedReadHeapLocFail
		\and
		\cmdRedAssignHeapLoc
		\and
		\cmdRedAssignHeapLocFail
		\and
		\cmdRedDesugarHeapCmdCall
	\end{mathpar}
	
	\caption{Command reduction rules.}
	\label{fig:def:CmdReductionRules}
\end{figure}

\section{Assertion Language}\label{sec:Assertions}

In the previous sections we introduced heap commands that capture the APIs of data structures.
In a similar way, we introduce heap predicates that describe their memory layout.

\begin{assumption}[Heap Predicates]
	We assume that there is a set of symbols \HeapPredSet.
	Every $\hPredVar \in \HeapPredSet$ represents a predicate characterising the heap.
	Further, we assume that there is a function 
	$\hpredInterpret : 
		\HeapPredSet 
		\rightarrow 
		\powerSetOf{\PhysHeapSet \times \ValueSet^*}$
	that maps each heap predicate symbol to a predicate over heaps and value tuples.
\end{assumption}

\begin{definition}[Assertions]
	We define the set of \emph{assertions} \AssertionSet according to the syntax presented in Figure~\ref{fig:def:Assertions}.
	
	We omit the index set \indexSetVar in quantifications when its choice becomes clear from the context and write 
	$\exists \indexVar.\, \assFam[\indexVar]$
	and
	$\forall \indexVar.\, \assFam[\indexVar]$
	instead of                
	$\exists \indexVar \in \indexSetVar.\, \assFam[\indexVar]$
	and
	$\forall \indexVar \in \indexSetVar.\, \assFam[\indexVar]$,
\end{definition}

\begin{figure}
	\begin{subfigure}{\textwidth}
		$$
		\begin{array}{c}
			\expVar \in \ExpSet
			\\
			\hPredVar \in \HeapPredSet
			\\
			\assSet \subseteq \AssertionSet
			\\
			\text{Index set}\ \indexSetVar \subseteq \Z
			\\
			\\
			\begin{array}{l c l}
				\assVar \in \AssertionSet
				&\ ::= \
				&\slTrue \alt
				\slFalse \alt
				\expVar \alt
				\neg \assVar \alt
				\assVar \wedge \assVar \alt
				\assVar \vee \assVar \alt
				\assVar \slStar \assVar \alt
				\assVar \slWand \assVar \alt
				\\
				&&\slPointsTo{\expVar}{\expVar} \alt 
				\hPredVar(\overline{\expVar}) \alt
				\bigvee \assSet \alt
				\slPersistent \assVar
			\end{array}
		\end{array}
		$$
		\caption{Assertion syntax.}
	\end{subfigure}

	\begin{subfigure}{\textwidth}
		$$
		\begin{array}{l c l}
		\assVar_1 \rightarrow \assVar_2
		\ \ &:= \ \
		&\neg\assVar_1 \vee \assVar_2
		\\
		\assVar_1 \leftrightarrow \assVar_2
		\ \ &:= \ \
		&(\assVar_1 \rightarrow \assVar_2) \wedge
		(\assVar_2 \rightarrow \assVar_1)
		\\
		\exists \indexVar \in \indexSetVar.\ \assFam[\indexVar]
		\ \ &:= \ \
		&\bigvee \setOf{\assFam[\indexVar] \ | \ \indexVar \in \indexSetVar}
		\\
		\forall \indexVar \in \indexSetVar.\ \assFam[\indexVar] 
		\ \ &:=\ \
		&\neg\exists \indexVar \in \indexSetVar.\, \neg \assFam[\indexVar]
		\end{array}
		$$
		\caption{Syntactic sugar.}
	\end{subfigure}
	
	\caption{Assertions.}
	\label{fig:def:Assertions}
\end{figure}

\begin{definition}{Assertion Model Relation}\label{def:AssertionModelRelation}
	We define the assertion model relation $\assModelsSymb\ \subseteq \PhysHeapSet \times \AssertionSet$ by recursion over the structure of assertions according to the rules presented in Fig.~\ref{fig:def:AssertionModelRelation}.
\end{definition}

\begin{figure}
	$$
	\begin{array}{l l l l}
	\assModels{\pheapVar}{\slTrue}
	\\
	\assNotModels{\pheapVar}{\slFalse}
	\\
	\assModels{\pheapVar}{\expVar}
		&\text{iff}
		&\assModels{\emptyset}{\evalExp{\expVar}}
		\\
	\assModels{\pheapVar}{\neg\assVar}
		&\text{iff}
		&\assNotModels{\pheapVar}{\assVar}
	\\
	\assModels{\pheapVar}{\assVar_1 \wedge \assVar_2}
		&\text{iff}
		&\assModels{\pheapVar}{\assVar_1}
			\wedge
			\assModels{\pheapVar}{\assVar_2}
	\\
	\assModels{\pheapVar}{\assVar_1 \vee \assVar_2}
		&\text{iff}
		&\assModels{\pheapVar}{\assVar_1}
			\vee
			\assModels{\pheapVar}{\assVar_2}
	\\
	\assModels{\pheapVar}{\assVar_1 \slStar \assVar_2}
		&\text{iff}
		&\exists \pheapVar_1, \pheapVar_2.\
			\pheapVar = \pheapVar_1 \disjCup \pheapVar_2
	\\
		&&\phantom{\exists \pheapVar_1, \pheapVar_2.\ }
			\wedge
			\assModels{\pheapVar_1}{\assVar_1}
			\wedge
			\assModels{\pheapVar_2}{\assVar_2}
	\\
	\assModels{\pheapVar}{\assVar_1 \slWand \assVar_2}
		&\text{iff}
		&\forall \pheapVar_1.\
			\pheapVar_1 \cap \pheapVar = \emptyset
			\wedge
			\assModels{\pheapVar_1}{\assVar_1}
	\\
		&&\phantom{\forall \pheapVar_1.\ }
			\rightarrow
			\assModels
				{\pheapVar_1 \disjCup \pheapVar}
				{\assVar_2}
	\\
	\assModels{\pheapVar}{\slPointsTo{\hlocVar}{\valVar}}
		&\text{iff}
		&\slPointsTo{\hlocVar}{\valVar} \in \pheapVar
	\\
	\assModels{\pheapVar}{\hPredVar(\expVar_1, \dots, \expVar_n)}
		&\text{iff}
		&\hpredInterpret[\hPredVar](\pheapVar, \evalExp{\expVar_1}, \dots, \evalExp{\expVar_n})
	\\
	\assModels{\pheapVar}{\bigvee \assSet}
		&\text{iff}
		&\exists \assVar \in \assSet.\	
			\assModels{\pheapVar}{\assVar}
	\\
	\assModels{\pheapVar}{\slPersistent \assVar}
		&\text{iff}
		&\assModels{\emptyset}{\assVar}
	\\
	\\
	\end{array}
	$$
	\caption{
		Assertion model relation.
		We write \assNotModels{\pheapVar}{\assVar} if \assModels{\pheapVar}{\assVar} does not hold.
	}
	\label{fig:def:AssertionModelRelation}
\end{figure}

\begin{definition}[Free Variables (Assertions)]\label{def:freeVariables_assertions}
	We define the notion of free variables for assertions analogously to that of commands (cf.~\ref{def:freeVariables_commands}).
\end{definition}

\begin{definition}[Substitution (Assertions)]\label{def:substitution_assertions}
	We define substitution for assertions analogously to substitution for commands (cf.~\ref{def:substitution_commands}).
\end{definition}

\begin{notation}{Free Variables of Tuples}
	For convenience we define the following notation for any tuple of commands and assertions
	$(y_1, \dots, y_n) \in (\CmdSet \cup \AssertionSet)^*$:
	$$
		\freeVars[y_1, \dots, y_n] 
		\ :=\
		\freeVars[y_1] \cup \dots \cup \freeVars[y_n].
	$$
\end{notation}

\begin{definition}[Validity]
	Let $\assVar \in \AssertionSet$ be an assertion with
	$\freeVars[\assVar] 
		\congruent
		\overline{\varVar}
		= (\varVar_i)_i
	$.
	For each $i = 1, \dots, n$, let $\typeVar_i$ be the type of variables $\varVar_i$ and let
	$\overline{\typeVar} = (\typeVar_i)_i$.
	We call assertion \assVar \emph{valid} if the following holds:
	$$
		\forall \pheapVar.\
		\forall \overline{\valVar} \in \overline{\typeVar}.\
		\assModels
			{\pheapVar}
			{\subst
				{\assVar}
				{\overline{\varVar}}
				{\overline{\valVar}}
			}
	$$
	We denote validity of \assVar by writing \assValid{\assVar}.
\end{definition}

\section{Memory Safety}\label{sec:MemorySafety}

\begin{definition}[Memory Safety of Commands and Heaps]
	We define the safety relation for commands 
	$\safeCmdHeapName \subseteq \PhysHeapSet \times \CmdSet$ 
	as follows:

	Let 
	$\overline{\varVar} = (\varVar_i)_i 
	\congruent
	\freeVars[\cmdVar]
	$
	be the variables occurring freely in \cmdVar.
	For each~$i$, let $\typeVar_i$ be the type of variable $\varVar_i$ and let $\overline{\typeVar} = (\typeVar_i)_i$.
	Then,
	$$
	\safeCmdHeap{\pheapVar}{\cmdVar}
	\ \Leftrightarrow \
	\forall \overline{\valVar} \in \overline{\typeVar}.\
	\neg\exists \nextPheapVar.\
		\cmdRedStepStar
			[\pheapVar]
			{\subst
				{\cmdVar}
				{\overline{\varVar}}
				{\overline{\valVar}}
			}
			[\nextPheapVar]
			{\memoryError}
	$$
	We say that a command \cmdVar is \emph{safe} under a physical heap \pheapVar if \safeCmdHeap{\pheapVar}{\cmdVar} holds.
\end{definition}

We consider a command safe under a heap if its execution does not lead to a memory error.
Note that a command's execution can get stuck without any memory error occurring.
Such cases arise for not-well-typed commands such as \cmdIfNoElse{13}{\dots}.
For this work, we only consider well-typed programs.
Hence, we do not care about cases in which a program gets stuck as long as no memory error occurs.

\begin{definition}[Memory Safety of Commands and Assertions]
	We define the safety relation for commands 
	$\safeCmdAssName \subseteq \AssertionSet \times \CmdSet$ 
	as follows:
	
	Let 
	$\overline{\varVar} = (\varVar_i)_i 
		\congruent
	 	\freeVars[\assVar, \cmdVar]
	 $
	be the variables occurring freely in \assVar and \cmdVar.
	For each~$i$, let $\typeVar_i$ be the type of variable $\varVar_i$ and let $\overline{\typeVar} = (\typeVar_i)_i$.
	Then,
	$$
		\safeCmdAss{\assVar}{\cmdVar}
		\quad \Longleftrightarrow \quad
		\forall \overline{\valVar} \in \overline{\typeVar}.\
		\forall \pheapVar.\
		\big(
			\assModels
				{\pheapVar}
				{\subst
						{\assVar}
						{\overline{\varVar}}
						{\overline{\valVar}}
				}
			\ \Rightarrow\
			\safeCmdHeap
				{\pheapVar}
				{\subst
						{\cmdVar}
						{\overline{\varVar}}
						{\overline{\valVar}}
				}
		\big)
	$$
	We say that a command \cmdVar is \emph{safe} under a assertion \assVar if \safeCmdAss{\assVar}{\cmdVar} holds.
\end{definition}

\begin{notation}
	We denote preconditions by \preVar.
	Further, we aggregate preconditions \preVar and programs \cmdVar into tuples \preProgTuple{\preVar}{\cmdVar}.
\end{notation}

\section{Verification Conditions}\label{sec:VerificationConditions}

\NewDocumentCommand{\symbCmdContextID}{}
	{\contextIdFont{scmds}}

\NewDocumentCommand{\symbEvalExp}{O{\cdot} O{\cdot} O{\cdot}}
	{\ensuremath{
		\evalExp{#1, #2, #3}
	}\xspace}

\NewDocumentCommand{\UsedVarSetVar}{}{\metaVar{U}}
\NewDocumentCommand{\symbStoreVar}{}{\metaVar{s}}
\NewDocumentCommand{\VcSetVar}{}{\metaVar{VC}}
\NewDocumentCommand{\vcVar}{}{\metaVar{vc}}

A common approach in program verification is to derive a verification condition \vcVar from the program \cmdVar and correctness property $\phi$ in question~\cite{Flanagan2001AvoidingExpExplosionVC, Parthasarathy2021VCGenerator}.
Instead of verifying the program directly, we prove the verification condition.
In general, \vcVar describes an over-approximation of all possible program behaviours.
The process is sound iff truth of the verification condition indeed implies that our program is correct, i.e., $\models \vcVar \Rightarrow \cmdVar \models \phi$.
We proceed analogously during our study of completeness thresholds.

\begin{definition}[Verification Condition]\label{def:verificationCondition}
	We call an assertion $\vcVar\in \AssertionSet$ a \emph{verification condition} for \preProgTuple{\preVar}{\cmdVar} if the following holds:
	$$
		\assValid{\vcVar}
		\quad\Rightarrow\quad
		\safeCmdAss{\preVar}{\cmdVar}
	$$
	We denote verification conditions by \vcVar.
\end{definition}

\begin{definition}[Precise Verification Conditions]\label{def:preciseVerificationCondition}
	Let \vcVar be a verification condition for \preProgTuple{\preVar}{\cmdVar} and let $\varVar \in \freeVars[\preVar, \cmdVar]$ be a free variable of type \typeVar.
	We call \vcVar 
	\emph{precise in \varVar for \preProgTuple{\preVar}{\cmdVar}}
	if the following holds for every value $\valVar \in \typeVar$:
	$$
		\safeCmdAss
			{\subst{\preVar}{\varVar}{\valVar}}
			{\subst{\cmdVar}{\varVar}{\valVar}}
		\quad\Rightarrow\quad
		\assValid{\subst{\vcVar}{\varVar}{\valVar}}
	$$
	
\end{definition}

\NewDocumentCommand{\arrayPred}{o}
{\ensuremath{
		\fixedPredNameFont{array}
		\IfValueT{#1}{(#1)}
	}\xspace}

\begin{assumption}[Array Predicate]
	We assume that there exists a predicate symbol $\arrayPred \in \HeapPredSet$.
	We further assume that 
	$\hpredInterpret[\arrayPred] \subseteq \PhysHeapSet \times \HeapObjSet \times \N$ 
	is the minimal relation for which the following holds:
	$$
	\begin{array}{l c l}
	\hpredInterpret[\arrayPred](\pheapVar, \arrayVar, \sizeVar)
	&\Longleftrightarrow
	&\displaystyle
	\assModels
	{\pheapVar}
	{
		\slBigStar_{0 \leq \indexVar < \sizeVar}
		\slPointsTo{(\arrayVar, \indexVar)}{\slWildcard}
	}
	\end{array}
	$$
	We identify the assertion \arrayPred[\arrayVar, \sizeVar] with the assertion 
	$\displaystyle
	\slBigStar_{0 \leq \indexVar < \sizeVar}
	\slPointsTo{(\arrayVar, \indexVar)}{\slWildcard}
	$.
\end{assumption}

\NewDocumentCommand{\postVar}{}{\metaVar{Q}}
\NewDocumentCommand{\resVar}{}{\metaVar{r}}
\NewDocumentCommand{\invVar}{}{\metaVar{I}}

\subsection{VC Generation}

\NewDocumentCommand{\AssertionLambdaSet}{}
{\metaVar{\AssertionSet_\lambda}}

\NewDocumentCommand{\lambdaIgnore}{m}
{\ensuremath{
		\lambda \_.\ #1
	}\xspace}

We use weakest liberal preconditions~\cite{Flanagan2001AvoidingExpExplosionVC, Dijsktra1976DisciplineOfProgramming} as verification conditions.

\begin{definition}{Assertion Lambdas}
	We define the set of assertion lambda terms as
	$\AssertionLambdaSet :=
	\setOf
	[\lambda \varVar.\ \assVar]
	{\varVar \in \VarSet, \assVar \in \AssertionSet}
	$.
	For convenience we define the notation
	$\lambdaIgnore{\assVar} := \lambda \varVar_{fresh}.\ \assVar$
	for $\varVar_{fresh} \not\in \freeVars[\assVar]$.
\end{definition}

We use assertion lambda terms $\lambda \resVar. \postVar$ to denote postconditions referring to a result value \resVar.

\begin{definition}{Weakest Liberal Precondition}
	We define the weakest liberal precondition function 
	$wlp : \CmdSet \times \AssertionLambdaSet \rightarrow \AssertionSet$
	by recursion over the structure of commands as follows:
	$$
	\begin{array}{l c l}
		wlp(\expVar,\ \lambda\resVar.\ \postVar)
		&:=
		&\subst{\postVar}{\resVar}{\expVar},
		\\
		wlp(
			\cmdLet{\varVar}{\cmdVar_1}{\cmdVar_2},\
			\lambda \resVar.\ \postVar
		)
		&:=
		&wlp(
			\cmdVar_1,\ 
			\lambda \varVar.\ 
				wlp(\cmdVar_2,\ \lambda \resVar.\ \postVar)
		),
		\\
		wlp(
			\cmdIf{\expVar}{\cmdVar_t}{\cmdVar_f},\
			\lambda \resVar.\ \postVar
		)
		&:=
		&(\expVar 
			\rightarrow 
			wlp(\cmdVar_t,\ \lambda \resVar.\ \postVar)
		)
		\\
		&&
		\wedge\
		(\neg\expVar 
			\rightarrow 
			wlp(\cmdVar_f,\ \lambda \resVar.\ \postVar)
		),
		\\
		wlp(\cmdWhileInv{\cmdReadHloc{\expVar}}{\invVar}{\cmdVar}, 
			\lambda \resVar.\ \postVar
		)
		&:=
		&\invVar
		\\
		&&\slStar\
		\slPersistent(
			\invVar \ \slWand\ \slPointsTo{\expVar}{\slWildcard}
		)
		\\
		&&\slStar\ 
		\slPersistent (
			\invVar \wedge \slPointsTo{\expVar}{\slTrue} 
			\ \slWand\
			wlp(\cmdVar, \lambdaIgnore{\invVar})
		)
		\\
		&&\slStar\ 
		(
			\invVar \wedge \slPointsTo{\expVar}{\slFalse} 
			\ \slWand\
			\postVar
		),
		\\
		wlp(\cmdForInv
			{\varVar}
			{\rangeFromTo{\expVar_1}{\expVar_2}}
			{\invVar}
			{\cmdVar},
		&:=
		&\invVar
		\\
		\phantom{wlp(}
		\lambda \resVar.\ \postVar)
		&&\slStar\
			\slPersistent(
				\forall \varVar.\ 
					\expVar_1 \leq \varVar < \expVar_2
					\wedge
					\invVar
					\ \slWand\
					wlp(\cmdVar, \lambdaIgnore{\invVar})
			)
		\\
		&&\slStar\
			(\invVar \ \slWand\ \postVar),
 		\\
		wlp(\cmdReadHloc{\expVar}, \lambda \resVar.\ \postVar)
		&:=
		&\exists y.\ 
			\slPointsTo{\expVar}{y}
			\wedge
			\subst{\postVar}{\resVar}{y}
		\\
		&&\text{where}\quad
		y \not\in \freeVars[\expVar, \postVar],
		\revRemark[Alternative wlp]
		{$
				\exists y.\
				\slPointsTo{\expVar}{y}
				\slStar
				(
					\slPointsTo{\expVar}{y}
					\slWand
					\subst{\postVar}{\resVar}{y}
				)
		$}
		\\
		wlp(\cmdAssignHloc{\expVar_1}{\expVar_2}, \lambda\resVar.\ \postVar)
		&:=
		&\slPointsTo{\expVar_1}{\slWildcard}
			\ \slStar\
			(
				\slPointsTo{\expVar_1}{\expVar_2}
				\slWand
				\postVar
			),
		\\
		wlp(\hCmdVar(\overline{\expVar}), \lambda\resVar.\ \postVar)
		&:=
		&wlp(
			\subst
				{\bCmdVar}
				{\overline{\varVar}}
				{\overline{\expVar}},
				\lambda \resVar.\ \postVar)
		\\
		&&\text{where}\quad
		\hcmdInterpret[\hCmdVar] = (\bCmdVar, \overline{\varVar})
	\end{array}
	$$
\end{definition}

Note that our weakest precondition for \keyword{for} loops does not allow the postcondition to depend on the knowledge that any work was done.
This simplified precondition is sufficient to reason about the memory safety of array-traversing programs that do not allocate nor free memory.

\deleteBlock{
\begin{example}[WLP]
	$$
	\begin{array}{l l}
		&wlp(
		\cmdLet
		{x}
		{(\cmdIf{a > 0}{a}{0})}
		{x},\
		\lambda r.\ r > 0
		)
		\\
		=
		&wlp(
		\cmdIf{a > 0}{a}{0},\
		\lambda x.\
		wlp(x,\ \lambda r.\ r > 0)
		)
		\\
		=
		&wlp(
		\cmdIf{a > 0}{a}{0},\
		\lambda x.\ x > 0
		)
		\\
		=
		&(a > 0 
		\ \rightarrow\
		wlp(a,\ \lambda x.\ x > 0)
		)
		\\
		&\wedge\
		(\neg(a > 0)
		\ \rightarrow\
		wlp(0,\ \lambda x.\ x > 0)
		)
		\\
		=
		&(a > 0  \ \rightarrow\ a > 0) \ \wedge\
		(\neg(a > 0) \ \rightarrow\ 0 > 0)
		)
		\\
		\equiv 
		&\slTrue \wedge (\neg(a > 0) \rightarrow \slFalse)
		\\
		\equiv 
		&a > 0
	\end{array}
	$$
\end{example}
}

\NewDocumentCommand{\hcmdVCs}{o}
{\ensuremath{
		vcs_\heapCmdContextID
		\IfValueT{#1}{(#1)}
	}\xspace}

\deleteBlock{
	\begin{assumption}[Verification Conditions for Heap Commands]
		We assume that there is a partial function 
		$\hcmdVCs : 
		\HeapCmdSet \times (\ValueSet \times \VarSet)^*
		\partialFunArrow
		\AssertionSet
		$
		that maps every well-typed heap command call to a verification condition.
	\end{assumption}
}

\deleteBlock{
	In the following definition, $\cdot[\cdot] \in \HeapCmdSet$ denotes the binary array access heap command.
	Normally, we write applications in infix notation, e.g., $\arrayVar[\indexVar]$.
}

\deleteBlock{
	\begin{assumption}[Verification Conditions for Arrays]
		We assume that \hcmdVCs maps array accesses to the following verification conditions:
		$$
		\begin{array}{l c l l}
			(\cdot[\cdot],\ \arrayVar, \indexVar)
			&\mapsto
			&\exists \sizeVar.\ 
			\arrayPred[\arrayVar, \indexVar] 
			\wedge
			0 \leq \indexVar \wedge \indexVar < \sizeVar
			\\
			(\cdot[\cdot],\ y_1, \dots, y_n)
			&\mapsto
			&\funUndefVal
			&\ \text{if}\quad
			n \neq 2
		\end{array}
		$$
	\end{assumption}
}

\deleteBlock{
	Note that according to the above definition \hcmdVCs is undefined for applications of the array access command to a wrong number of arguments.
}

\deleteBlock{
	\begin{definition}{Relaxed Liberal Precondition}
		We define the relaxed liberal precondition function 
		$rlp : \CmdSet \times \AssertionLambdaSet \rightarrow \AssertionSet$
		analogous to $wlp$ but with the adjustment that heap commands are handled by \hcmdVCs.
	\end{definition}
}

\section{Completeness Thresholds}\label{sec:CompletenessThresholds}

\NewDocumentCommand{\ctVar}{}{\metaVar{R}}
\NewDocumentCommand{\ctElemVar}{}{\metaVar{r}}

Now that we have an intuition for what a completeness threshold should be and for how we want to use it, let's formalise this intuition.

\begin{definition}[Completeness Thresholds for Quantified Assertions]\label{def:ass:completenessThreshold}
	Let $Q \in \setOf{\forall, \exists}$ be a quantifier and let 
	$Q\varVar \in \typeVar.\ \assVar$ be a quantified assertion.
	Further, let $\ctVar \subseteq \typeVar$ be a restriction of the domain of \varVar.
	We call \ctVar a \emph{completeness threshold} for \assVar if the following holds:
	$$
		\assValid{
			Q \varVar \in \ctVar.\ \assVar
		}
		\quad\Rightarrow\quad
		\assValid{
			Q \varVar \in \typeVar.\ \assVar
		}.
	$$
\end{definition}

\begin{definition}[Completeness Thresholds for Programs]\label{def:prog:completenessThreshold}
	Let \typeVar be the type of \varVar in \preProgTuple{\preVar}{\cmdVar} and let $\ctVar \subseteq \typeVar$ be a restriction of this type.
	We call \ctVar a \emph{completeness threshold} for \varVar in \preProgTuple{\preVar}{\cmdVar} if the following holds:
	$$
		\forall \valVar \in \ctVar.\
		\safeCmdAss
			{\subst{\preVar}{\varVar}{\valVar}}
			{\subst{\cmdVar}{\varVar}{\valVar}}
		\quad\Rightarrow\quad
		\forall \valVar \in \typeVar.\
		\safeCmdAss
			{\subst{\preVar}{\varVar}{\valVar}}
			{\subst{\cmdVar}{\varVar}{\valVar}}
	$$
\end{definition}

\begin{lemma}[Precision]
	Let $\forall \varVar \in \typeVar.\ \vcVar$ be a verification condition for \preProgTuple{\preVar}{\cmdVar} and let $\ctVar \subseteq \typeVar$ be a completeness threshold for \vcVar.
	Further, let \vcVar be precise in \varVar for \preProgTuple{\preVar}{\cmdVar}.
	Then \ctVar is a completeness threshold for \preProgTuple{\preVar}{\cmdVar}.
\end{lemma}
\begin{proof}
	Assume that
	$\forall \valVar \in \ctVar.\ 
		\safeCmdAss
			{\subst{\preVar}{\varVar}{\valVar}}
			{\subst{\cmdVar}{\varVar}{\valVar}}
	$
	holds.
	We have to prove that the program is safe for the unrestricted domain, i.e., \safeCmdAss{\preVar}{\cmdVar}.
	
	Together with the precision of \vcVar in \varVar, this assumption implies
	$\forall \valVar \in \ctVar.\ 
		\assValid{
			\subst{\vcVar}{\varVar}{\valVar}
		}
	$
	(cf. Def.~\ref{def:preciseVerificationCondition}),
	which is equivalent to
	$\assValid{
			\forall \varVar \in \ctVar.\ 
			\vcVar
		}
	$.
	
	\ctVar is a completeness threshold for \vcVar.
	According to Def.~\ref{def:ass:completenessThreshold}, this means
	$\assValid{
			\forall \varVar \in \typeVar.\ 
			\vcVar
		}
	$.
	Since the latter is a verification condition for \preProgTuple{\preVar}{\cmdVar}, 
	proposition
	\safeCmdAss{\preVar}{\cmdVar}
	holds by Def.~\ref{def:verificationCondition}.
	Hence, \ctVar is a completeness threshold for \preProgTuple{\preVar}{\cmdVar}.
\end{proof}

Finding completeness threshold for a non-precise verification condition $\forall \varVar.\ \vcVar$ does not allow us to conclude that we found a completeness threshold for the actual program \preProgTuple{\preVar}{\cmdVar}.
However, as long as our ultimate goal is to verify the program by proving an assertion $A$ that is at least as strong as our verification condition, i.e., $A \Rightarrow \forall \varVar.\ \vcVar$,
we can leverage the completeness threshold.
Hence, it makes sense to first concentrate on completeness thresholds for verification conditions.
Later, we can try to relate our results to completeness thresholds for programs.

\begin{lemma}\label{lemma:assWithRestrictedDomain}
	Let $\forall \varVar \in \typeVar.\ \assVar$ be an assertion.
	Let $\ctVar \subseteq \typeVar$ be such that
	$$
	\forall \valVar_1, \valVar_2 \in \ctVar.\quad
		(
			\assValid{\subst{\assVar}{\varVar}{\valVar_1}}
			\ \ \Leftrightarrow\ \
			\assValid{\subst{\assVar}{\varVar}{\valVar_2}}
		).
	$$
	Then, for every $\ctElemVar \in \ctVar$ it holds that
	$$
		\assValid{
			\forall \varVar \in 
				(\typeVar\setminus\ctVar) \cup \setOf{\ctElemVar}.\
				\assVar
		}
		\quad\Leftrightarrow\quad
		\assValid{
			\forall \varVar \in \typeVar.\
			\assVar
		}.
	$$
\end{lemma}
\begin{proof}
	Let $\ctElemVar \in \ctVar$.
	With the assumption from the lemma, we get
	$$
		\forall \valVar \in \ctVar.\quad
		(
			\assValid{\subst{\assVar}{\varVar}{\ctElemVar}}
			\ \ \Leftrightarrow\ \
			\assValid{\subst{\assVar}{\varVar}{\valVar}}
		)
	$$
	and hence
	$$
		\assValid{\subst{\assVar}{\varVar}{\ctElemVar}}
		\ \ \Leftrightarrow\ \
		\assValid{\forall \varVar \in \ctVar.\ \assVar}
	$$
	Further,
	$\forall \varVar \in \typeVar.\ \assVar$
	is valid iff both
	$\forall \varVar \in (\typeVar \setminus \ctVar).\ \assVar$
	and
	$\forall \varVar \in \ctVar.\ \assVar$
	are valid.
	As we can reduce validity of the latter to validity of \subst{\assVar}{\varVar}{\ctElemVar},
	we get
	$$
	\begin{array}{l l}
			&(\assValid
			{
				\forall \varVar \in 
					(\typeVar \setminus \ctVar).\ 
					\assVar
			})
			\ \wedge \
			(\assValid
			{
				\subst{\assVar}{\varVar}{\ctElemVar}
			})
		\\
		\Longleftrightarrow
			&(\assValid
			{
				\forall \varVar \in 
				(\typeVar \setminus \ctVar).\ 
				\assVar
			})
			\ \wedge \
			(\assValid{ \forall \varVar \in \ctVar.\ \assVar})
		\\
		\Longleftrightarrow
			&\assValid{\forall \varVar \in \typeVar.\ \assVar}
		.
	\end{array}
	$$

\end{proof}

\begin{corollary}\label{corollary:ass:domainRestrictionCT}
	Let $\forall \varVar \in \typeVar.\ \assVar$ be an assertion.
	Let $\ctVar \subseteq \typeVar$ be such that
	$$
		\forall \valVar_1, \valVar_2 \in \ctVar.\quad
		(
			\assValid{\subst{\assVar}{\varVar}{\valVar_1}}
			\ \ \Leftrightarrow\ \
			\assValid{\subst{\assVar}{\varVar}{\valVar_2}}
		).
	$$
	Then, for every $\ctElemVar \in \ctVar$, the set 
	$(\typeVar \setminus \ctVar) \cup \setOf{\ctElemVar}$
	is a completeness threshold for assertion \assVar.
\end{corollary}
\begin{proof}
	Follows from Lem.~\ref{lemma:assWithRestrictedDomain} and Def.~\ref{def:ass:completenessThreshold}.
\end{proof}

Consider a verification condition $\forall \varVar \in \typeVar.\ \vcVar$ and suppose we are interested in a completeness threshold for \varVar.
By definition, the threshold is a restriction of \varVar's domain, i.e., $\ctVar \subseteq \typeVar$.
The lemmas and corollary above show us that one way forward is to identify a validity-preserving subset of \typeVar.
That is, we need to look for a subset $\ctVar \subseteq \typeVar$ of the domain within which the concrete choice for \varVar does not affect the validity of the verification condition.
Once we got this, we can collapse \ctVar to any representative $\ctElemVar \in \ctVar$ and we found our completeness threshold $(\typeVar \setminus \ctVar) \cup \setOf{\ctElemVar}$.

Notice that validity preservation of domain restrictions is a transitive property.
Hence, we can easily turn the search for completeness thresholds for a fixed variable into an iterative approach.
\begin{lemma}[Transitivity of Completeness Thresholds for Fixed Variable]\label{lemma:CT_Transitivity_fixedVar}
	Let $\ctVar_0, \ctVar_1, \ctVar_2$ be sets with $\ctVar_2 \subseteq \ctVar_1 \subseteq \ctVar_0$.
	Let $\assVar_i = \forall \varVar \in \ctVar_i.\ \assVar$ be assertions.
	Let $\ctVar_1$ and $\ctVar_2$ be completeness thresholds for $\varVar$ in $\assVar_0$ and $\assVar_1$, respectively.
	Then, $\ctVar_2$ is also a completeness threshold for \varVar in $\assVar_0$.
\end{lemma}
\begin{proof}
	Since $\ctVar_1$ is a completeness threshold for \varVar in 
	$\assVar_0 = \forall \varVar \in \ctVar_0.\ \assVar$, 
	we get
	$$
		\assValid{
			\forall \varVar \in \ctVar_1.\ \assVar
		}
		\quad\Rightarrow\quad
		\assValid{
			\forall \varVar \in \ctVar_0.\ \assVar
		}.
	$$
	Since $\ctVar_2$ is a completeness threshold for \varVar in 
	$\assVar_1 = \forall \varVar \in \ctVar_1.\ \assVar$, 
	we get
	$$
		\assValid{
			\forall \varVar \in \ctVar_2.\ \assVar
		}
		\quad\Rightarrow\quad
		\assValid{
			\forall \varVar \in \ctVar_1.\ \assVar
		}.
	$$
	That is,
	$$
	\assValid{
		\forall \varVar \in \ctVar_2.\ \assVar
	}
	\quad\Rightarrow\quad
	\assValid{
		\forall \varVar \in \ctVar_1.\ \assVar
	}
	\quad\Rightarrow\quad
	\assValid{
		\forall \varVar \in \ctVar_0.\ \assVar
	}
	$$
	and hence $\ctVar_2 \subset \ctVar_0$ is a completeness threshold for \varVar in
	$\assVar_0 = \forall \varVar \in \ctVar_0.\ \assVar$
\end{proof}

\begin{corollary}
	Let \typeVar be a type and let $(\ctVar_i)_i$ be a family of sets with $\ctVar_0 = \typeVar$ and $\ctVar_{i+1} \subseteq \ctVar_i$.
	Let $(\assVar_i)_i = (\forall \varVar \in \ctVar_i.\ \assVar)_i$ be a family of assertions such that each $\ctVar_{i+1}$ is a completeness threshold for \varVar in $\assVar_i$.
	Then, each $\ctVar_i$ is a completeness threshold for \varVar in 
	$\assVar_0 = \forall \varVar \in \typeVar.\ \assVar$.
\end{corollary}
\begin{proof}
	Follows from Lem.~\ref{lemma:CT_Transitivity_fixedVar} by induction.
\end{proof}

Consider a program that traverses an array $a$ of size $\sizeVar_a$ and an array $b$ of size $\sizeVar_b$.
When we analyse the verification condition of this program we find that it contains distinct parts that describe memory safety of the accesses to array $a$ and distinct parts for the accesses to $b$.
Since both arrays describe separate parts of the heap, we can bring the verification condition into a form that reflects this.
Thereby, we get a formula of the form
$\vcVar \equiv \vcVar_a \slStar \vcVar_b$
where $\vcVar_a$ and $\vcVar_b$ describe memory safety in respect to $a$ and $b$, respectively.

Suppose, we want to find a completeness threshold for $\sizeVar_a$.
In some cases, the manipulation of both arrays is entangled which means that $\sizeVar_a$ potentially affects the validity of $\vcVar_b$.
In such a case, we have no choice but to analyse the entire formula to find our completeness threshold.
However, often that's not the case and $\sizeVar_a$ only shows up in the subformula $\vcVar_a$ that actually concerns array $a$.
In this case, it is sufficient to analyse $\vcVar_a$ in order to find a completeness threshold for $\sizeVar_a$.

\begin{lemma}[Elimination]\label{lemma:Elimination}
	Let $\assVar, \assVar_x, \assVar'$ be assertions with
	$\forall \varVar \in \typeVar.\ \assVar
		\equiv
		\forall \varVar \in \typeVar.\ \assVar_\varVar \slStar \assVar'
	$.
	Suppose the choice of \varVar does not affect the validity of $\assVar'$, i.e.,
	$$
		\forall \valVar \in \typeVar.\quad
		(
			\assValid{\assVar'}
			\quad\Leftrightarrow\quad
			\assValid{\subst{\assVar'}{\varVar}{\valVar}}
		).
	$$
	Let $\ctVar \subseteq \typeVar$ be a completeness threshold for \varVar in
	$\forall \varVar \in \typeVar.\ \assVar_x$.
	Then, \ctVar is also a completeness threshold for \varVar in 
	$\forall \varVar \in \typeVar.\ \assVar$.
\end{lemma}
\begin{proof}
	Since $\ctVar$ is a completeness threshold for \varVar in 
	$\forall \varVar \in \ctVar.\ \assVar_x$, 
	we get
	$$
	\begin{array}{l l}
		&\assValid{
			\forall \varVar \in \ctVar.\ \assVar_x \slStar \assVar'
		}
		\\
		\quad\Rightarrow\quad
		&\assValid{
			(\forall \varVar \in \ctVar.\ \assVar_x)
			\slStar
			(\forall \varVar \in \ctVar.\ \assVar')
		}
		\\
		\quad\Rightarrow\quad
		&\assValid{
			(\forall \varVar \in \typeVar.\ \assVar_x)
			\slStar
			(\forall \varVar \in \ctVar.\ \assVar')
		}.
	\end{array}
	$$
	By using the assumption that the choice of \varVar does not affect the validity of $\assVar'$ we can conclude
	$$
	\begin{array}{l l}
		&\assValid{
			(\forall \varVar \in \typeVar.\ \assVar_x)
			\slStar
			(\forall \varVar \in \ctVar.\ \assVar')
		}
		\\
		\quad\Rightarrow\quad
		&\assValid{
			(\forall \varVar \in \typeVar.\ \assVar_x)
			\slStar
			(\forall \varVar \in \typeVar.\ \assVar')
		}		
		\\
		\quad\Rightarrow\quad
		&\assValid{
			\forall \varVar \in \typeVar.\ \assVar_x \slStar \assVar'
		}.
	\end{array}
	$$
\end{proof}

\begin{corollary}[VC Slicing]
	Let $\assVar, \assVar_x, \assVar'$ be assertions with
	$\forall \varVar \in \typeVar.\ \assVar
	\equiv
	\forall \varVar \in \typeVar.\ \assVar_\varVar \slStar \assVar'
	$.
	Suppose \varVar is not free in $\assVar'$.
	Let $\ctVar \subseteq \typeVar$ be a completeness threshold for \varVar in
	$\forall \varVar \in \typeVar.\ \assVar_x$.
	Then, \ctVar is also a completeness threshold for \varVar in 
	$\forall \varVar \in \typeVar.\ \assVar$.
\end{corollary}
\begin{proof}
	Follows from Lem.~\ref{lemma:Elimination}.
\end{proof}

\subsection{Iteratively Extracting Completeness Thresholds}

What we saw so far, gives us the tools to define an iterative process to extract completeness thresholds.

\begin{workflow}
	Let \preProgTuple{\preVar}{\cmdVar} be a program with variables $\varVar_1, \dots, \varVar_n$ for which we would like to extract completeness thresholds.
	\begin{enumerate}
		\item
			Compute a verification condition for \preProgTuple{\preVar}{\cmdVar}, e.g., by using weakest preconditions.
			The result has the form
			$\forall \overline{\varVar} \in \overline{\typeVar}.\ \vcVar$.
		\item 
			Iteratively extract completeness thresholds for each $\varVar_i$.\\
			For all $i \in \setOf{1, \dots, n}$:
			\begin{enumerate}
				\item 
					Let $\overline{\ctVar} \subseteq \overline{\typeVar}$ be the completeness thresholds extracted so far.
					(Initially $\overline{\ctVar} = \overline{\typeVar}$.)
				\item
					Bring the verification condition into the form\\
					$\forall \overline{\varVar} \in \overline{\ctVar}.\ \vcVar
						\quad\equiv\quad
						\forall \overline{\varVar} \in \overline{\ctVar}.\ 
							\slBigStar_{0 \leq j \leq m} \vcVar_j
					$.
				\item 
					Identify a subformula $\vcVar'$ whose validity is not affected by the choice of $\varVar_i$.
					Bring the verification condition into the form\\
					$\forall \overline{\varVar} \in \overline{\ctVar}.\ \vcVar
						\quad\equiv\quad
						\forall \overline{\varVar} \in \overline{\ctVar}.\ 
							\vcVar_i \slStar \vcVar'
					$.\\
					In the remaining steps it suffices to analyse
					$\forall \overline{\varVar}\ \in \overline{\ctVar}.\ \vcVar_i$.
				\item
					Examine $\vcVar_i$ and extract a completeness threshold.
					This can either be done purely manually or by identifying patterns for which we previously proved that we can extract completeness thresholds.\\
					This step yields a completeness threshold $\ctVar_i'$ which, in the worst case, does not yield an improvement, i.e., $\ctVar_i' = \ctVar_i$.
				\item 
					Repeat this process iteratively until the extracted completeness threshold does not improve in respect to the last iteration.
			\end{enumerate}
	\end{enumerate}
\end{workflow}

\subsection{Iterating over Arrays}

\NewDocumentCommand{\rangeLeft}{}{\metaVar{a}}
\NewDocumentCommand{\rangeRight}{}{\metaVar{b}}

In the following, we study patterns encountered in verification conditions of programs that iterate over arrays.
The goal of this section is to formulate reusable lemmas that allow us to automate the extraction of completeness thresholds.

\begin{lemma}[VC-CT for Bounded, Unconditional Array Access]
	\label{lemma:vc:ct:boundedUnconditionalArrayAccess}
	Let $\zVar, \rangeLeft, \rangeRight \in \Z$ be constants and
	$\vcVar(\sizeVar) = 
		\forall \indexVar \in \Z.\
			\rangeLeft \leq \indexVar < \sizeVar+\rangeRight
			\rightarrow
			0 \leq \indexVar + \zVar < \sizeVar
	$.
	Then, for every $\ctElemVar \in \Z$ with $\ctElemVar > \rangeLeft - \rangeRight $ it holds that
	$$
	\assValid{
		\forall \sizeVar \in \Z.\  \vcVar(\sizeVar)
	}
	\quad\Leftrightarrow\quad
	\assValid{\vcVar(\ctElemVar)}
	$$
	That is, \setOf{\ctElemVar} is a completeness threshold for $\vcVar(s)$.
\end{lemma}
\begin{proof}
	For $\sizeVar \leq \rangeLeft-\rangeRight$ it holds
	$\rangeLeft \leq \indexVar < \sizeVar+\rangeRight \equiv \slFalse$
	and hence
	$\vcVar(\sizeVar) \equiv \slTrue$.
	
	For $\sizeVar > \rangeLeft - \rangeRight$ we get:
	$$
	\begin{array}{r l l}
		\vcVar(\sizeVar) \equiv
			&\forall \indexVar.\
				\rangeLeft \leq \indexVar < \sizeVar + \rangeRight
				\rightarrow
				0 \leq \indexVar + \zVar < \sizeVar
		\\
		\equiv
			&\forall \indexVar.\
				(
					\rangeLeft \leq \indexVar < \sizeVar + \rangeRight
					\rightarrow
					0 \leq \indexVar + \zVar
				)
				\wedge
				(
					\rangeLeft \leq \indexVar < \sizeVar + \rangeRight
					\rightarrow
					\indexVar + \zVar < \sizeVar			
				)				
		\\
		\equiv
			&\forall \indexVar.\
			(
				\rangeLeft \leq \indexVar
				\rightarrow
				0 \leq \indexVar + \zVar
			)
			\wedge
			(
				\indexVar < \sizeVar + \rangeRight
				\rightarrow
				\indexVar + \zVar < \sizeVar			
			)
		\\
		\equiv
			&\forall \indexVar.\
			(
				\rangeLeft \leq \indexVar
				\rightarrow
				0 \leq \indexVar + \zVar
			)
			\wedge
			\forall \indexVar.\
			(
				\indexVar < \sizeVar + \rangeRight
				\rightarrow
				\indexVar + \zVar < \sizeVar
			)
		\\
		\equiv
			&\forall \indexVar.\
			(
				\rangeLeft \leq \indexVar
				\rightarrow
				0 \leq \indexVar + \zVar
			)
			\wedge
			\forall \indexVar.\
			(
				\indexVar < \rangeRight
				\rightarrow
				\indexVar + \zVar < 0
			)
		\\
		=: 
			&\vcVar^+
	\end{array}
	$$
	\sizeVar does not occur in $\vcVar^+$ and, hence, the concrete value of \sizeVar does not impact the validity of $\vcVar^+$.
	By Lem.~\ref{corollary:ass:domainRestrictionCT},
	for any choice of $\ctElemVar \in \Z$ with $\ctElemVar > l -r $, the set \setOf{\ctElemVar} is a completeness threshold for $\vcVar(s)$.
\end{proof}

\section{Roadmap}\label{sec:Roadmap}
	This report describes a work in progress.
	We are currently studying patterns encountered in programs that traverse arrays.
	Once sufficiently many patterns have been identified, we are going to prove that we can extract completeness thresholds for a relevant class of array-traversing programs.
	Afterwards, we are going to generalize our findings to arbitrary tree-like inductive data structures.

\bibliographystyle{plain} 
\bibliography{bibliography}

\end{document}